# Memristor Model Based on Fuzzy Window Function

Rabab Farouk Abdel-Kader[1], Sherif M. Abuelenin[2]
Electrical Engineering Department
Faculty of Engineering, Port-Said University
Port-Said, Egypt
[1]rababfakader@eng.psu.edu.eg, [2]s.abuelenin@eng.psu.edu.eg

*Abstract*—**Memristor (memory-resistor) is the fourth passive circuit element. We introduce a memristor model based on a fuzzy logic window function. Fuzzy models are flexible, which enables the capture of the pinched hysteresis behavior of the memristor. The introduced fuzzy model avoids common problems associated with window-function based memristor models, such as the terminal state problem, and the symmetry issues. The model captures the memristor behavior with a simple rule-base which gives an insight of how memristors work. Because of the flexibility offered by the fuzzy system, shape and distribution of input and output membership functions can be tuned to capture the behavior of various real memristors.**

## I. INTRODUCTION

Memristors (contraction for memory–resistors), the fourth fundamental passive circuit element, were theorized in 1970's [1]. Although memristive behavior was reported in several publications [1]-[6], it was not until 2008 when HP announced the manufacturing of the first Titanium Dioxide (*TiO₂*) practical memristor [2]. Later, other realizations were introduced. Following the development of real memristors, several applications were introduced that benefit from their behavior. This includes analog, digital, and memory applications.

Memristors are characterized by a pinched hysterics *I-V* curve. Unlike the theoretical memristors, real ones exhibit complex behavior characterized by complex shaped hysterics curves. In order to capture the complex behavior of practical memristors accurately, several models (with different math complexity and accuracy) were introduced. Memristor models can be categorized into three categories [12];

1. Liner dopant drift assumption modeling [3]; where the memristor is considered to be acting as two variable resistors connected in series and their values change in response to excitation field, which causes carriers migration. The drift velocity of carriers is assumed to be constant throughout the entire length of the memristor, hence the name, linear drift models. Also there is no mathematical account for the finite limits of physical boundaries. This is the simplest and least accurate representation, as it fails to accurately describe the physical behavior.

2. Models based on nonlinear dopant drift assumptions (e.g. [4]-[9]). A window function is added to impose a changing drift velocity and resolve the boundary conditions. Window function based nonlinear models are discussed in more details in the following section.

3. Tunneling based modeling [10], [11]; where the memristor is modeled as a tunnel barrier sandwiched between two contacts. This approach shows closer agreement with the HP *TiO₂* memristor data. Also, recent research [15] suggested the coexistence of both dopant drift and tunneling effects.

In this paper we introduce a fuzzy logic window function based model. The model accurately captures the memristors behavior, and its rule base provides an insight of how memristors work.

The rest of this paper is organized as follows. The next section lists and briefly explains different window function based linear and non-linear dopant drift models. Fuzzy-logic-based window function is introduced in the following section, simulation and results are then discussed, and finally conclusion and future work are provided in the last section.

## II. WINDOW FUNCTION MODELS

Among the different memristive devices modeling approaches listed in previous section, we limit the discussion here to window function based models.

A diagram of the memristor is shown in Fig. 1, along with its equivalent circuit model. The mathematical model suggested [3] for the resistance of a memristor can be described by (1) and (2).

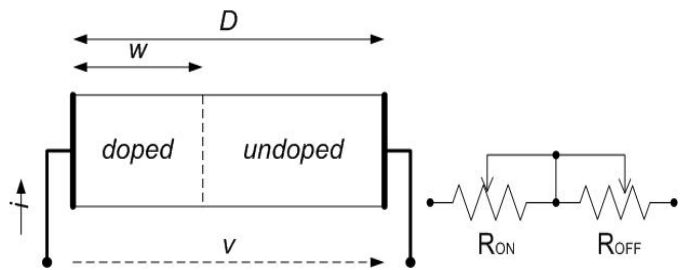

Fig. 1. Memristor diagram and its equivalent model [3]

$$R(w) = R_{on} \cdot \frac{w}{D} + R_{off} \cdot (1 - \frac{w}{D}) \quad (1)$$

$$\frac{dw(t)}{dt} = \mu_v \cdot \frac{R_{on}}{D} \cdot i(t) \quad (2)$$

where $\mu_v$ is the average ion mobility, and $w(t)$ is the width of the doped region. $R_{on}$ and $R_{off}$ are the minimum and maximum resistance values attained when the whole width of the memristor is either doped or un-doped, respectively. Equation (2) shows that the width of the doped region is proportional to the integral of the current.

This model is based on linear doping drift, and is invalid at boundaries where ($w < 0$ or $w > D$). In nano-scale devices, small voltages can yield enormous electric fields [3], which can produce significant nonlinearities in ionic transport.

The utilization of a window function $f(x)$ was introduced to account for the nonlinear dopant drift effect [3]. The window function is multiplied by the right-hand side of (2) to model the behavior at the boundaries, as well as other nonlinear dynamic effects of the memristive device behavior.

Sturkov et al. [3] suggested the window function $f(x) = w(1-w)/D^2$, which corresponds to nonlinear drift when $w$ is close to either boundary (x = zero or x = D). The equation can be written in terms of the state variable as

$$f(x) = x - x^2 \quad (3)$$

The boundary condition at the *OFF-state*, i.e., when $w = 0$, is resolved since $f(x) = 0$. The window function also imposed nonlinear behavior over the device. However, it suffers from the terminal-state problem. i.e., when $w$ reaches one of the terminal values $(x = 0$ or $x = 1)$, the state of the device cannot be changed any further. This can be seen from (2), when the value of the window function becomes zero, the derivative of the state variable equals zero, and the state variable of the model maintains its current value. This means that when setting the memristive device to the terminal state $R_{ON}$ or $R_{OFF}$, no external stimulus can change this state, i.e. such a memristor would be bound to hold its state forever [8].

Prodromakis et al. [9] listed the following conditions for an effective window function;

1) Take into account the boundary conditions at the top and bottom electrodes of the device;
2) Be capable of imposing nonlinear drift over the entire active core of the device;
3) Provide linkage between the linear and nonlinear dopant drift models;
4) Be scalable, meaning a range of $f_{max}(x)$ can be obtained such that $0 \leq f_{max}(x) \leq 1$;
5) Utilize a built-in control parameter for adjusting the model.

Joglekar and Wolf [5] proposed an alternative window function:

$$f(x) = 1 - (2x - 1)^{2p} \quad (4)$$

It still suffered the same terminal-state problem associated with (3).

Biolek et al. [8] suggested the following window function to enhance the mathematical representation of the memristor;

$$f(x) = 1 - (x - \text{sgn}(-i))^{2p} \quad (5)$$

where, $sgn(i) = 1$ for $i \geq 0$, and $sgn(i) = 0$ for $i < 0$. This window function resolved the fact that the boundary speeds of approaching and receding from the thin film limits are different. However, it lacked scalability by not having a parameter to control the magnitude of the function.

In order to address the five conditions listed earlier for an effective window function, Prodromakis et al. [9] proposed the following form;

$$f(x) = j\{1 - [(x - 0.5)^2 + 0.75]^p\} \quad (6)$$

where $p$ and $j$ (both $\in R+$) are two controlling parameters. The parameter '$p$' determines the rate of decrease of the window function as the state variable approaches any of its two bounds, and '$j$' sets the maximum value of the window function itself, which may here exceed the unitary value [13].

The four above discussed window functions are of various complexities, and show various degrees of accuracies in modeling memristors characteristics. Reference [13] provides a comparison among them, it is shown that neither Joglekar's nor Prodomakis' models were able to reproduce the dynamics shown by the reference model. It is noted that all of them are functions only in the state variable (except Biolek's, which is also function of the sign, not the magnitude, of the current). Therefore, none of these window functions takes into account the threshold effect (the minimum voltage [7], [15], or current [16], that is needed for the memristance effect to appear). If the applied voltage (or device current) is below the threshold level, no noticeable change in memristance can appear.

It is also noted that all of the discussed window functions are symmetric around $x=0.5$ (Except for Biolek's window function, which is symmetric around $x = 0.5$ for opposite signs of the device current). This can cause them to be incapable of reproducing certain dynamics.

The BCM [4] model was introduce to provide more accuracy, utilizing a parametric window function. This model is based on a window function having unitary value for all values of $x(t) \in (0, 1)$, and exhibiting vertical transitions for certain cases. The model is rather accurate, but has several parameters, and represented using a set of rules. Also, regardless of its good accuracy, it does not satisfy some of the window function conditions discussed earlier.

Fuzzy modeling is an extension of rule-base modeling, whose linguistic rule-base provides an understanding of how the system dynamics work. Its utilization in modeling memristive systems is discussed in the next section.

### III. FUZZY WINDOW FUNCTION

We utilize the flexibility of fuzzy modeling to introduce a fuzzy window function that is capable of addressing the aforementioned issues. It also gives insights of how real memristors behave, by describing the operation in a set of linguistic *if-then* rules.

Fig. 2(a) shows the membership functions for the input *I*, Fig. 2(b) shows the membership function for the input *x*, and Fig. 2(c) shows the membership function for the output *F*. The rule-bases are shown in Fig. 2(d), and the surface of the window function is shown in Fig. 2(e). The '*min*' operator is used for '*and*' and for implication, '*max*' is used for aggregation, and centroid defuzzification is utilized.

These show one of the possible window functions. Problems of boundary conditions and requirement of symmetry of the behavior for positive and negative currents are resolved. In general, the fuzzy window function is flexible to tune, making it satisfying all of the previously discussed conditions for an effective window function.

### IV. SIMULATION AND RESULTS

Simulations are performed using discrete modeling in Matlab®. Memristor parameters used in the simulations are:

$k=10000$, $R_{on}=100\ \Omega$, $R_{off}=16\ k\Omega$, $R_{init}=11\ k\Omega$.
where $k=mu*R_{on}/D^2$

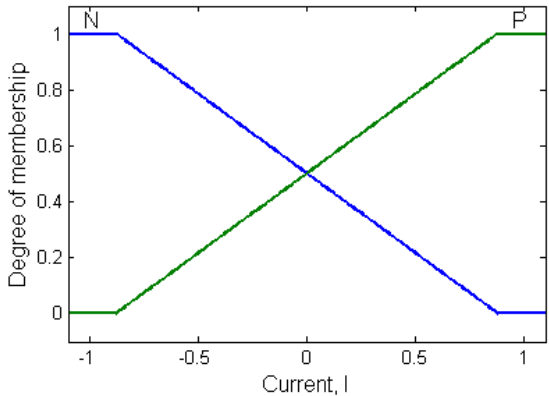

Fig. 2(a). Membership functions for the input *I* (memristor current), note that the x axis is scaled by a factor of 1000.

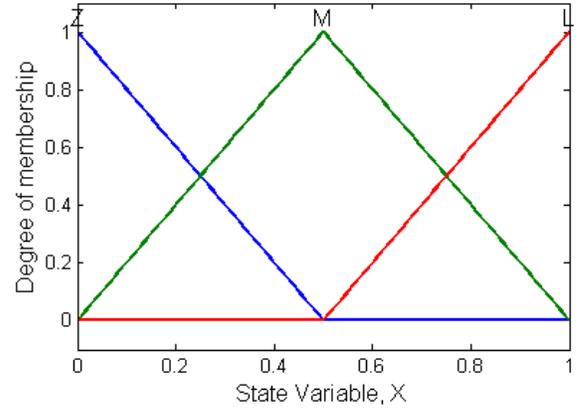

Fig. 2(b). Membership functions for the input *x* (state variable, *x=w/D*)

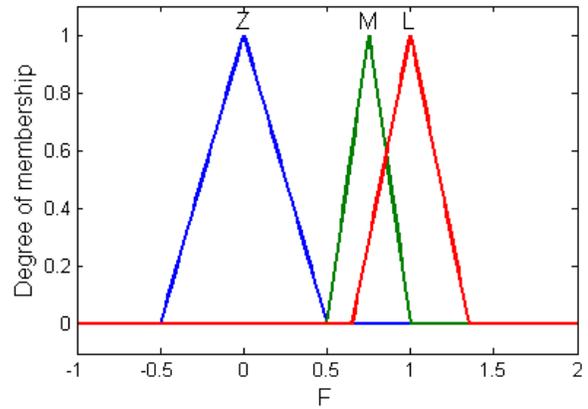

Fig. 2(c). Membership functions for the output F (window function value)

1. If (I is N) and (X is Z) then (F is Z) (1)
2. If (I is N) and (X is M) then (F is M) (1)
3. If (I is N) and (X is L) then (F is L) (1)
4. If (I is P) and (X is Z) then (F is L) (1)
5. If (I is P) and (X is M) then (F is M) (1)
6. If (I is P) and (X is L) then (F is Z) (1)

Fig. 2(d). Rule-base

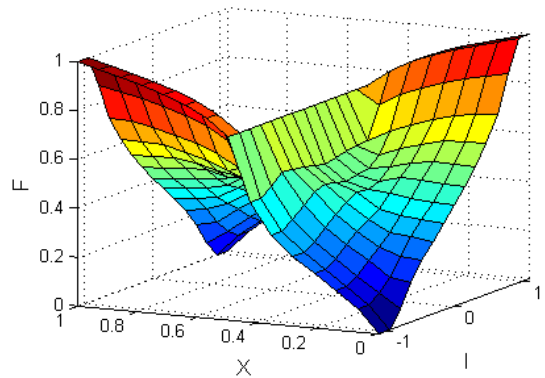

Fig. 2(e). Fuzzy window-function surface

The results introduced here are using a circuit in which the memristor is placed in series with a *2 kΩ* resistor and a voltage source. Fig. 3(a) shows the memristor *I-V* curve for one period of a large input voltage signal ($v(t) = 5\sin(\omega t)$). The signal drives the memristor to its limit. Both signals as a function of time are shown in Fig. 3(b). When the state space in Joglekar's window function (with *p=10*) model reaches its maximum value, it maintains this value, causing the memristor to keep a constant value while the current is changing represented as a red-dashed curve. Clearly this is not the case with the fuzzy model as shown with the blue-solid curve.

One additional aspect to consider in modeling memristors is the threshold required for the memristance effect to appear (the migration of dopants). The fuzzy system can alternatively be made a function of the memristor voltage instead of its current. This allows setting a voltage threshold instead of current threshold. Fig. 4 shows one possible window function, that takes into account a voltage threshold. This is done by adding a third (Zero) membership function to the input space of the Current (or Voltage) linguistic variable. This membership function is chosen to be trapezoidal. The rule base is modified to include a 7$^{th}$ rule;

*If V is Z then F is Z.*

It is clear from the figure that the value of the function is zero (causing the derivative function to be zero, hence, no change in memristance) for voltage values less than the positive threshold and more than the negative threshold values (in our case, the magnitude of each is set to be around *0.2 V*). The values of both thresholds don't necessarily have to be equal. Fig. 5 shows the effect of threshold voltage, the circuit is excited using a sinusoidal voltage with peak value of *0.2 V*, which is equal to the threshold value.

Simulating the memristor that utilizes a threshold-based window function with large input voltages shows more non-linear effect (as expected). This can be seen in Fig. 6.

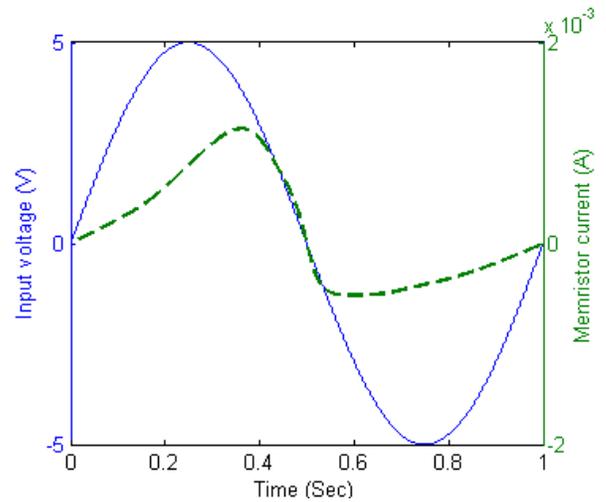

Fig. 3(b). Memristor *I* and *V* vs. time.

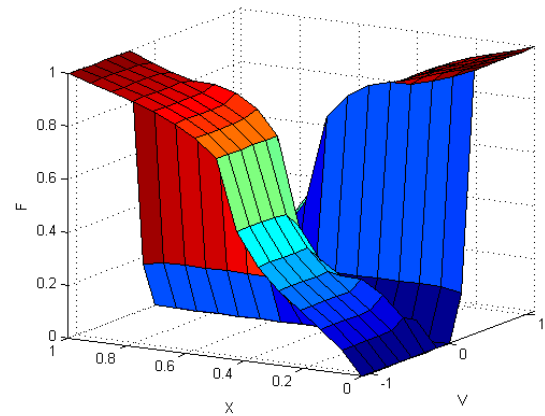

Fig. 4. Fuzzy window function with threshold

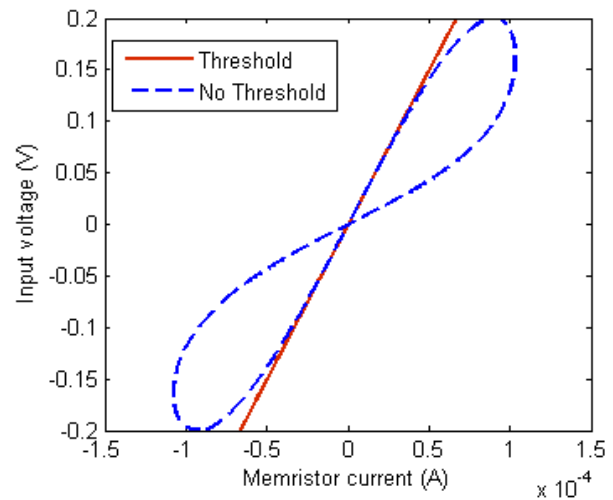

Fig. 5. Memristive I-V curve for two window functions; threshold and no-threshold considered.

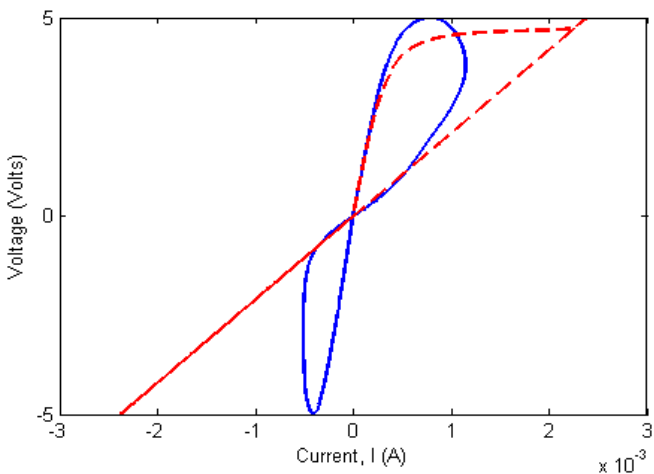

Fig. 3(a). Memristive *I-V* curve for *V(t)=5sin(ωt)*

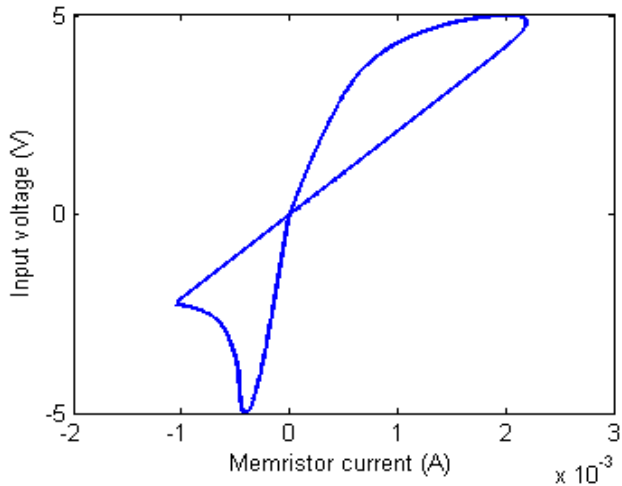

Fig. 6. Memristive *I-V* curve for threshold-based window functions with large input voltage.

V. CONCLUSION

We introduced a memristor model based on a fuzzy logic window function. The model was able to capture the pinched hysteresis behavior of the memristor. It permits the avoidance of common problems associated with window-function based memristor models, namely; boundary condition requirements, and the symmetry problem. The model captures the memristor behavior with a simple rule-base which gives an insight of how memristors work. Because of the flexibility offered by the fuzzy system, shape and distribution of input and output membership functions can be tuned to capture the behavior of various real memristors. This includes accounting for an excitation threshold. Future work includes tuning the fuzzy system to match the response of real fabricated memristors, and studying the effect of non-volatility.